\documentclass[twocolumn,showpacs,preprintnumbers,nofootinbib,prl,superscriptaddress,groupedaddress,10pt]{revtex4-1}

\usepackage{amssymb,amsmath,verbatim,mathtools,needspace,enumitem,etoolbox,graphicx}
\usepackage[dvipsnames, usenames]{xcolor}

\definecolor{linkcolor}{rgb}{0.0,0.3,0.5}
\usepackage[unicode, colorlinks=true, linkcolor=linkcolor, citecolor=linkcolor, filecolor=linkcolor,urlcolor=linkcolor, pdfusetitle]{hyperref}
\usepackage{epstopdf}

\usepackage{bm}
\usepackage{dcolumn}
\usepackage[utf8]{inputenc} 
\usepackage{latexsym}
\usepackage{rotating}
\usepackage{xcolor}
\usepackage{longtable}
\usepackage{enumerate}
\usepackage{mathtools}
\usepackage{url}
\renewcommand{\vec}[1]{\boldsymbol{#1}}

\newcommand{\ben}{\begin{enumerate}}
\newcommand{\een}{\end{enumerate}}

\def\be{\begin{equation}}
\def\ee{\end{equation}}
\def\bea{\begin{eqnarray}}
\def\eea{\end{eqnarray}}
\newcommand{\beq}{\begin{eqnarray}}
\newcommand{\eeq}{\end{eqnarray}} 
\newcommand{\ba}{\begin{align}}
\newcommand{\ea}{\end{align}}

\def\ba{\bar{a}}

\newcommand{\tn}{\textnormal}

\begin{document}
\title{Constraining the neutron star equation of state using multi-band independent measurements of radii and tidal deformabilities}
\author{Margherita Fasano}
\email{margherita.fasano@roma1.infn.it}
\affiliation{Dipartimento di Fisica, Sapienza Università di Roma \& Sezione INFN Roma1, P.A. Moro 5, 00185, Roma, Italy}

\author{Tiziano Abdelsalhin}
\email{tiziano.abdelsalhin@roma1.infn.it}
\affiliation{Dipartimento di Fisica, Sapienza Università di Roma \& Sezione INFN Roma1, P.A. Moro 5, 00185, Roma, Italy}

\author{Andrea Maselli}
\email{andrea.maselli@roma1.infn.it}
\affiliation{Dipartimento di Fisica, Sapienza Università di Roma \& Sezione INFN Roma1, P.A. Moro 5, 00185, Roma, Italy}

\author{Valeria Ferrari}
\email{valeria.ferrari@uniroma1.it}
\affiliation{Dipartimento di Fisica, Sapienza Università di Roma \& Sezione INFN Roma1, P.A. Moro 5, 00185, Roma, Italy}

\begin{abstract}
Using a Bayesian approach, we combine measurements of neutron star
macroscopic observables obtained by astrophysical and gravitational
observations, to derive joint constraints on the equation of state
(EoS) of matter at supranuclear density.  In our analysis we use  two
sets of data: (i) the masses and  tidal deformabilities measured in
the binary neutron star event GW170817, detected by LIGO and Virgo;
(ii) the masses and stellar radii measured from observations of
nuclear bursts in accreting low-mass X-ray binaries. Using a phenomenological 
parametrization of the equation of state, we compute the posterior
probability distributions of the EoS parameters, using which we infer
the posterior distribution for the radius and the mass of the two
neutron stars of GW170817.  The constraints we set on the radii are
tighter than previous bounds.  \end{abstract}
\maketitle

\noindent{\bf{\em Introduction.}}
The detection of the gravitational wave (GW) signal emitted in the
coalescence of the binary neutron star (BNS),
GW170817~\cite{TheLIGOScientific:2017qsa},  offers a unique
opportunity to probe the properties of matter at the extreme densities
occurring in a neutron star (NS) core.
This detection has stimulated a number of studies in this direction,
both by the LIGO/Virgo collaboration (LVC) and by independent groups.
As a result, new constraints on NS internal composition have been
derived, according to which  the  equation of state (EoS) of nuclear
matter is more likely to be soft, leading to stellar
configurations with high compactness and small
radii~\cite{Shibata:2017xdx,Rezzolla:2017aly,Margalit:2017dij,Ruiz:2017due,Most:2018hfd,Annala:2017llu,Bauswein:2017vtn,Coughlin:2018miv,Radice:2018ozg,Radice:2017lry,Abbott:2018wiz,Abbott:2018exr,De:2018uhw,Coughlin:2018fis,De:2018uhw,Carson:2018xri,Raithel:2018ncd,Zhang:2018vrx,Landry:2018prl,Malik:2018zcf,Tews:2018chv,Lim:2018bkq,Kumar2019}.

The GW signal emitted in a BNS coalescence
carries the imprint of NS structure in both the
inspiral and merger/post-merger
phases~\cite{Hinderer:2009ca,Maselli:2013rza,Read:2013zra,DelPozzo:2013ala,Wade:2014vqa,Bauswein:2011tp,Bose:2017jvk}.
During the inspiral, the information on the stellar composition is
encoded at the leading order in the quadrupolar
tidal deformability $\Lambda$, which describes how the shape of one star changes
in response to the external tidal field~\cite{Hinderer:2007mb}. For a
given equation of state, this parameter depends solely on the stellar
compactness ${\cal C}={M/R}$, i.e., on  the ratio between mass and
circumferential radius of the NS at equilibrium. Moreover, $\Lambda$
is a monotonic function of ${\cal C}$, and decreases as the
compactness grows, i.e. for more compressible (soft) matter. 

The analysis of the data of GW170817 has allowed to estimate
the average parameter\footnote{$\tilde{\Lambda}$ is the actual parameter
that enters at the leading order into the gravitational
waveform~\cite{Flanagan:2007ix,Vines:2010ca,Vines:2011ud}.}
\begin{equation}
\tilde{\Lambda}=\frac{16}{13}\frac{(M_1+12M_2)M_1^4\Lambda_1+(M_2+12M_1)M_2^4\Lambda_2}{(M_1+M_2)^5}\ ,
\end{equation}
where $\Lambda_{1,2}$ are the individual NS tidal deformabilities and 
$M_{1,2}$ are the NS masses~\cite{Abbott:2018wiz,Abbott:2018exr}.  Assuming low-spin
priors, the analysis carried out by the LVC collaboration yields a
value of $\tilde \Lambda=300^{+420}_{-230}$ at 90\% confidence level \cite{Abbott:2018wiz}.
Combined with the posterior distributions of the inferred masses, this
result leads to a constraint on the NS radius of $10 \lesssim R\lesssim 13$ km,
which excludes EoS predicting stiff matter,
i.e., less compact stars~\cite{Abbott:2018exr}.

The first BNS event also marks the dawn of  multi-messenger astronomy,
which will combine observations in the electromagnetic (EM) and in the
gravitational bandwidths,  expected to cover  a
broad range of wavelengths, and different stages of the evolution of
the observed sources.

Following the discovery of the EM counterpart of GW170817, the
information obtained from the GW data has been complemented  by that
inferred from the electromagnetic observations associated with the
merger/post-merger phase of the binary event~\cite{Monitor:2017mdv,GBM:2017lvd,Coulter:2017wya}; in particular,
the properties of the gamma ray burst and the kilonova light
curves have been exploited~\cite{Rezzolla:2017aly,Margalit:2017dij,Coughlin:2018fis,Radice:2018ozg,Radice:2017lry,Bauswein:2017vtn,Coughlin:2018miv,Most:2018hfd}.
Thus, in these studies EM and GW observations of the {\it same} event
have been fully exploited.

In this paper we combine {\it independent} measurements
of NS macroscopic observables obtained from EM and GW data
to derive joint constraints on the EoS of
matter at supranuclear density.   
We consider two distinct datasets 
based on: (i) masses and tidal deformabilities extracted from the the
data analysis of GW170817; (ii) masses and radii measured through
spectroscopic observations of NS thermonuclear bursts in low-mass
X-ray binaries~\cite{Nattila2016,Nattila2017}. Using a phenomenological 
parametrized EoS, we infer the posterior probability
distribution of the EoS parameters through a fully Bayesian analysis,
and derive new bounds on the radius of the two NS coalescing in
GW170817.  Our results show how our understanding of the EoS of matter
at supranuclear density can benefit from the synergy of data coming from
astrophysical phenomena spanning very different dynamical regimes, and
detected with distinct experimental setups.

Furthermore, the proposed approach is a promising tool to exploit the data of
high-precision surveys which, in the near future, will be available
from space satellites \cite{2014SPIE.9144E..20A} and from advanced and
third generation GW interferometers
\cite{ac03c906d03248f3a6a5229c15aa770d,2010CQGra..27a5003H}.

\noindent{\bf{\em Parametrized EoS.}}
The thermodynamical properties of matter inside a cold neutron star
can be described by a barotropic relation between the pressure $p$ and
the energy density $\epsilon$, i.e. by the equation of state,
$p=p(\epsilon)$. At densities  $\rho \lesssim \rho_0$, where $\rho_0
\sim 2.7 \times 10^{14} \mathrm{g/cm^3}$ is the equilibrium density of
nuclear matter, the EoS has been determined by extrapolating the
results of terrestrial  experiments on atomic
nuclei~\cite{Haensel:1993zw,Abrahamyan:2012gp,Tsang:2008fd,Tamii:2011pv},
and there is a general consensus on its properties. For densities
above the saturation point, typical of a NS core, the EoS is less
certain; indeed, due to the complexity of quantum chromo-dynamics and
to the difficulty of testing these regimes with experiments on Earth,
hadronic interactions are described by a variety of models based on
different approaches and assumptions~\cite{Lattimer:2000nx}. 
These EoS, when used to describe a NS, lead to different values of the 
observables and to different relations between radius, or tidal deformability, and
mass.  Thus, astrophysical and gravitational wave measurements of these 
observables can be exploited to constrain the EoS, solving the so-called 
\emph{inverse stellar problem}~\cite{Ozel:2009da,Read:2009yp,Steiner:2012xt,Lindblom:2012zi,Lindblom:2013kra,Lackey:2014fwa,Guillot:2014lla,Ozel:2015fia,Raithel:2017ity,Abdelsalhin:2017cih,Carney:2018sdv,Lindblom:2018ntw}.
Phenomenologically parametrized EoS are particularly
useful in this respect, as they allow us to describe a large class of
theoretical  EoS through a relatively small set of coefficients~\cite{Read:2008iy,Lindblom:2010bb,Steiner:2010fz,Raithel:2016bux,Lindblom:2018rfr}.
Moreover, they provide a unique tool to combine stellar parameters 
obtained from NS observations in different waveband, to infer features of the
\emph{true} EoS which may not be predicted by current models.  

In this work we focus on the spectral representation developed by Lindblom~\cite{Lindblom:2010bb}. 
This model is based on a series expansion of the adiabatic index $\Gamma(p)$. It has
been shown that most theoretical EoS are well approximated 
including the first four terms in the expansion, which correspond
to the 4 free parameters of the model $(\gamma_0,\gamma_1,\gamma_2,\gamma_3)$, 
such that 
\begin{equation}
 \Gamma(p) \simeq\exp \Biggl\{ \sum_{k=0}^{3}\gamma_k \left[\log\biggl(\frac{p}{p_0}\biggr)\right]^k\Biggl\} ,\label{spectral}
\end{equation}
where $p_0$ is the pressure at the crust-core interface, which we 
choose as in \cite{Abbott:2018exr}. 

\noindent{\bf{\em The Bayesian framework.}}
The goal of this work is to combine measurements of NS observables
obtained in different astrophysical channels, such as masses,
radii and tidal deformabilities, and to reconstruct the parameters of
a phenomenological EoS. The mass and the radius of a NS can be
measured through electromagnetic observations in the X-, optical
and radio wavebands of low-mass binary systems~\cite{Ozel:2016oaf, Nattila2016,Nattila2017}.  
The detection of gravitational waves emitted in BNS merging allows
to estimate masses and tidal deformabilities of the coalescing
bodies~\cite{TheLIGOScientific:2017qsa,Abbott:2018wiz,Abbott:2018exr,LIGOScientific:2018mvr}.
We shall now show how, using a Bayesian scheme of inference,
 such complementary information can be combined to put
stronger constraints on the NS EoS.

Within the spectral parametrization, the equilibrium configuration of a $i=1,\ldots n$ 
NSs is completely specified by the $m+n$ parameters, namely by the  $m=4$ coefficients 
$\gamma_k$ introduced in eq.~\eqref{spectral}, and and by $n$ values of the central pressure $p_{i=1,\ldots n}^c$.
On the other hand, each NS observation provides $2$ observables, either the mass 
and the radius $(M,R)$, or the mass and the tidal deformability $(M,\Lambda)$, obtained 
in the EM and GW channel, respectively. Therefore, if we want to fully characterise the 
EoS, and infer all the spectral parameters, we need at least $N=4$ observations, yielding
$2N=8$ macroscopic observables. This is the minimum number of data which would allow to 
determine the 8 unknown quantities
\begin{equation} 
\vec{\theta} = 
 ( \gamma_0,\gamma_1,\gamma_2,\gamma_3,
p^c_{1},p^c_{2},p^c_{3},p^c_{4})  \label{params} \ .
\end{equation}
Hereafter $p^c_{1,2}$ ($p^c_{3,4}$) will correspond to the pressures of 2 
NSs observed in the GW (EM) channel.

The LIGO/Virgo collaboration has detected one binary NS merger so far,
GW170817, and the mass and tidal
deformabilities of the two stars have been
estimated~\cite{Abbott:2018wiz,Abbott:2018exr}.  Estimates of NS masses and radii 
based on electromagnetic observations have been obtained using a wide variety of 
methodologies applied to different astrophysical environments. In this work we 
consider the studies carried out by N\"attil\"a and collaborators, in which the NS masses 
and radii are reconstructed by using the cooling tail method for the low-mass X-ray binary 
4U 1724-307~\cite{Nattila2016} and by fitting the X-ray bursting NSs directly to the observed 
spectra of 4U 1702-429~\cite{Nattila2017}. GW170817 and two
EM measurements provide enough information to solve the inverse
stellar problem described before, with a joint set of data given by 
$\vec{d} =\{(M_1,\Lambda_1,M_2,\Lambda_2), (M_3,R_3),(M_4,R_4)\}$.

Using a Bayesian approach, we compute the posterior probability
density function (PDF) of the EoS parameters given the experimental
data, $\mathcal{P}(\vec{\theta}|\vec{d})\propto  \mathcal{L}(\vec{d} |
\vec{\theta}) \mathcal{P}_0(\vec{\theta})$,
where $\mathcal{P}_0(\vec{\theta})$ is the prior on the parameters,
and $ \mathcal{L}(\vec{d} | \vec{\theta})$ is the likelihood function. The latter, 
in our case, reads:
\begin{align}
\mathcal{L}(\vec{d} | \vec{\theta}) = & \mathcal{L}^\tn{GW}(M_1,\Lambda_1,M_2,\Lambda_2) \times \nonumber\\
& \mathcal{L}^\tn{EM}(M_3,R_3) \times \mathcal{L}^\tn{EM}(M_4,R_4) \,,\label{likelihood}
\end{align}
where $\mathcal{L}^{GW}(M_1,\Lambda_1,M_2,\Lambda_2)$ and
$\mathcal{L}^{EM}(M_i,R_i)_{i=3,4}$ are the probability computed by
the LIGO/Virgo collaboration~\cite{Com:EoS,Ind:EoS} and and N\"attil\"a \emph{et
  al.}~\cite{Nattila2016,Nattila2017}, respectively.
We sample the posterior distribution using Markov chain Monte Carlo
(MCMC) simulations based on the Metropolis-Hastings algorithm
\cite{Gilks:1996}. The MCMC convergence is enhanced by a Gaussian
adaptation algorithm~\cite{1085030,5586491} (see
\cite{Abdelsalhin:2017cih} and reference therein for a detailed
discussion on this approach).

\noindent{\bf{\em Numerical setup.}}
For the GW data, the likelihood
$\mathcal{L}^{GW}(M_1,\Lambda_1,M_2,\Lambda_2)$ is given by the joint
probability distribution for the masses and tidal deformabilities  of
GW170817 inferred by the LVC using a parametrised EoS~\cite{Abbott:2018exr}. 
For the EM
sector, $\mathcal{L}^{EM}(M_i,R_i)_{i=3,4}$ is given by  the joint
probability distributions of the most accurate  measurements of masses
and radii provided in~\cite{Nattila2016, Nattila2017}. The latter correspond to
two NSs in low-mass X-ray binaries observed during thermonuclear
bursts, namely 4U 1724-307~\cite{Nattila2016} and 4U 1702-429~\cite{Nattila2017}.
Table~\ref{tab:Likelihood} shows the median and the 90\% confidence
intervals for the GW and EM data.  
\begin{table}[ht]
	\renewcommand*{\arraystretch}{1.2}
  \centering
     \begin{tabular}{cccc}
   \hline
   \hline
   $M_1~[M_\odot]$ & $\Lambda_1$ & $M_2~[M_\odot]$ & $\Lambda_2$ \\
 \hline
$1.46^{+0.13}_{-0.09}$ & $255_{-171}^{+416}$
&  $1.26_{-0.12}^{+0.09}$ & $661_{-375}^{+858}$ \\
\hline
 \hline
\\
  \hline
   \hline
      $M_3~[M_\odot]$ & $R_3$ [km] & $M_4~[M_\odot]$ & $R_4$ [km] \\
 \hline
$1.79_{-0.30}^{+0.20}$ & 
$12.81_{-0.87}^{+0.44}$ & 
$1.48_{-0.87}^{+0.44}$ & 
$11.06_{-2.06}^{+1.57}$\\
   \hline
   \hline
  \end{tabular}  
  \caption{Median and 90\% intervals for the masses $M_{1,2}$
and the tidal deformabilities $\Lambda_{1,2}$ of the two NS observed in
GW170817~\cite{Abbott:2018exr} and for the masses $M_{3,4}$ and the
radii $R_{3,4}$ of 4U 1702-429~\cite{Nattila2017} and 4U 1724-307~\cite{Nattila2016}.}
\label{tab:Likelihood} \end{table}

We assume a flat prior distributions for all parameters, namely: 
$\gamma_0 \in [0.2,2]$, $\gamma_1 \in [-1.6,1.7]$, $\gamma_2
\in [-0.6,0.6]$ and  $\gamma_3 \in [-0.02,0.02]$, in agreement with
the analysis made in Ref.~\cite{Abbott:2018exr}.  Furthermore, for
each combination of parameters, we require the corresponding adiabatic
index to be $\Gamma(p)\leq 7$ up to the central pressure of the NS
maximum mass configuration. Finally, we ask all central pressures to
be uniformly distributed in the range $p^c_i \in[10^{33},10^{38}]$
dyn/cm$^2$. Also, we ask that the inferred EoS is consistent with
existent astrophysical and theoretical bounds. Specifically, we
require that: (i) each EoS model supports a maximum mass $M_{max} \geq
1.97 M_{\odot}$~\cite{Antoniadis:2013pzd}, and (ii) the EoS remains
causal (speed of sound $c_s \leq c$, where $c$ is the speed of light
in vacuum) up to the maximum mass configuration.  For both models, we
run 24 independent chains of $n=2 \cdot 10^7$ samples, discarding the first
$\sim 10\%$ points as burn-in. The convergence of the MCMC is
determined by monitoring each chain autocorrelation function and
cross-checking the chains through a standard Rubin
test~\cite{Gilks:1996}.

\begin{figure}[!htbp]
\centering
\includegraphics[width=0.23\textwidth]{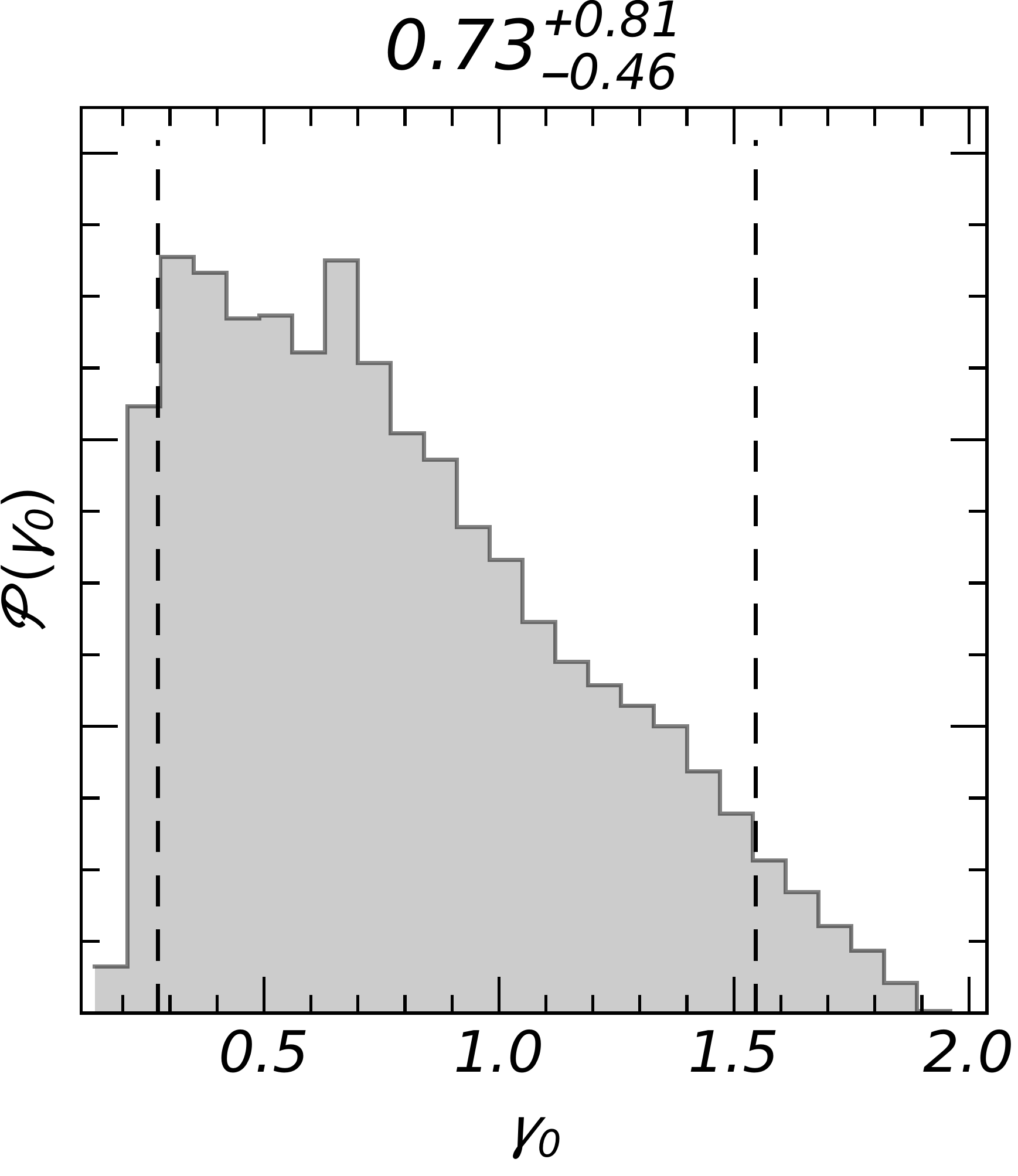}
\includegraphics[width=0.23\textwidth]{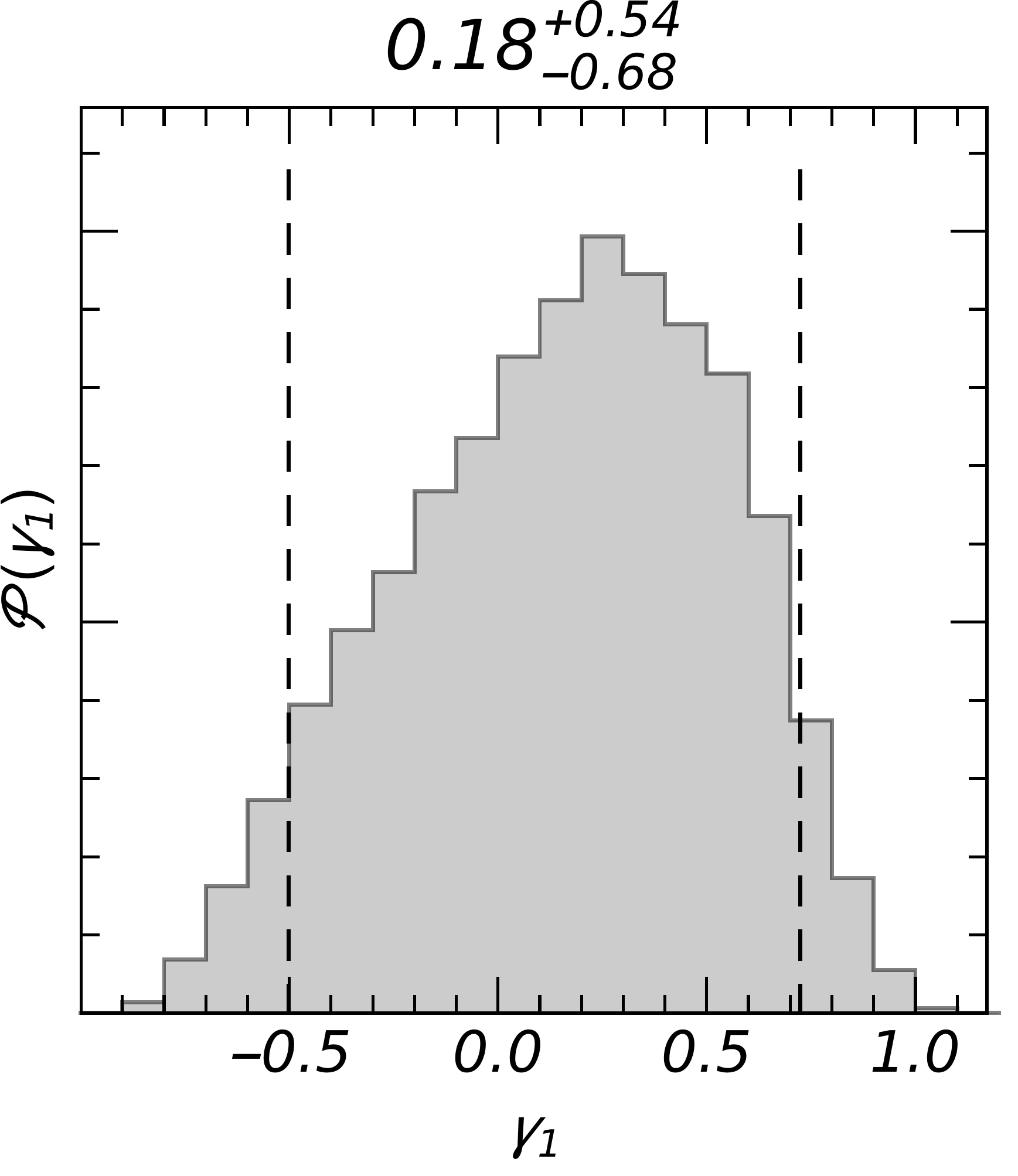}\\
\includegraphics[width=0.23\textwidth]{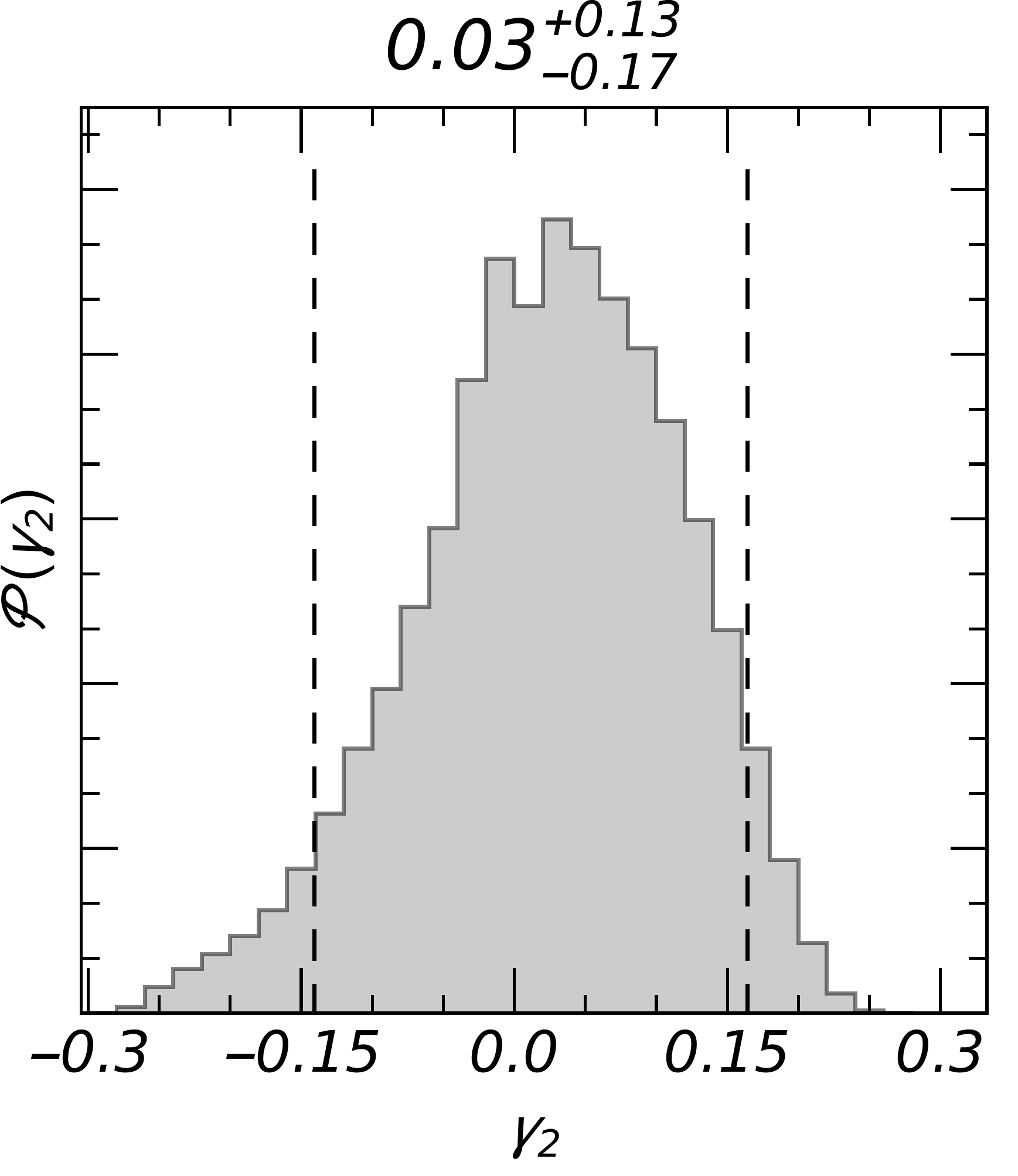}
\includegraphics[width=0.23\textwidth]{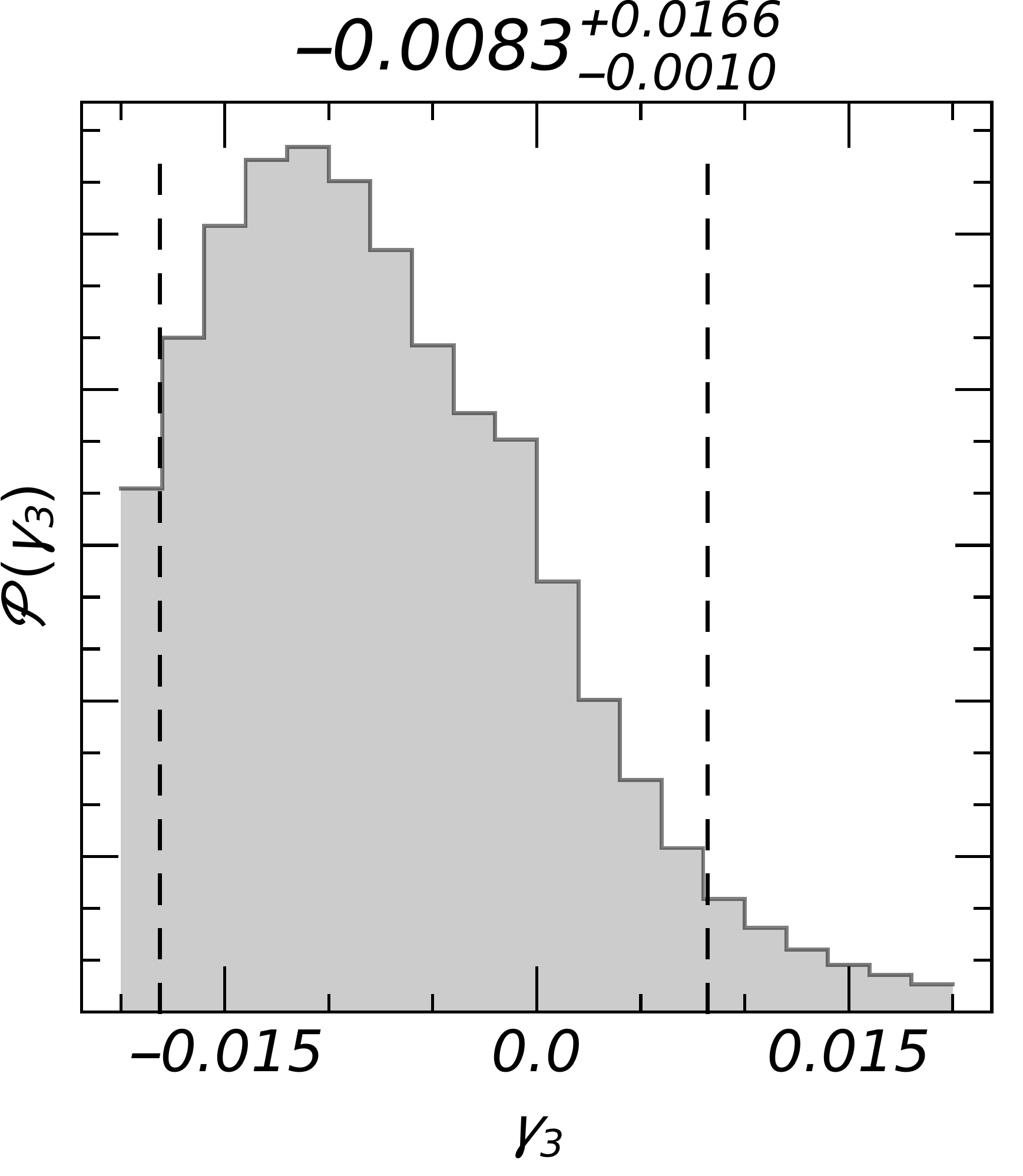}
\caption{Posterior probability distributions of the parameters 
of the spectral representations derived through the multi-messenger 
analysis. Dashed vertical lines identify 90\%   confidence intervals, 
also shown on top of each panel with the median.}
\label{fig:PWposteriors} 
\end{figure}

\noindent{\bf{\em Results.}}
The four panels of Fig~\ref{fig:PWposteriors} show the posterior
distributions for the parameters of the spectral equation of state. 
In each panel, dashed vertical lines identify 90\%
confidence intervals.
The lowest order term in the expansion of the adiabatic index $\Gamma(p)$, i.e. $\gamma_{0}$, 
provides the weakest bounds with respect to the priors, while the linear, $\gamma_1$, and the 
``high-density'', $\gamma_{2}$ and $\gamma_3$, coefficients lead to the strongest constraints. 
Such posteriors are consistent with previous studies on large density regimes within NS cores 
featuring quark matter \cite{Most:2018hfd,Annala:2017llu}. 
As indicated in (\ref{params}), the central pressures are also sampled in the MCMC.  
We find that for the neutron stars of GW170817 these quantities are in agreement with the values quoted by 
the LVC \cite{Abbott:2018exr}. \\

Having inferred the probability distributions
for the parameters \eqref{params},
we can compute the posteriors  for radius and mass
of the two NS of GW170817, by solving the TOV equations
for each of the model sampled by the MCMC.

\begin{figure}[!htbp] \centering
\includegraphics[width=0.35\textwidth]{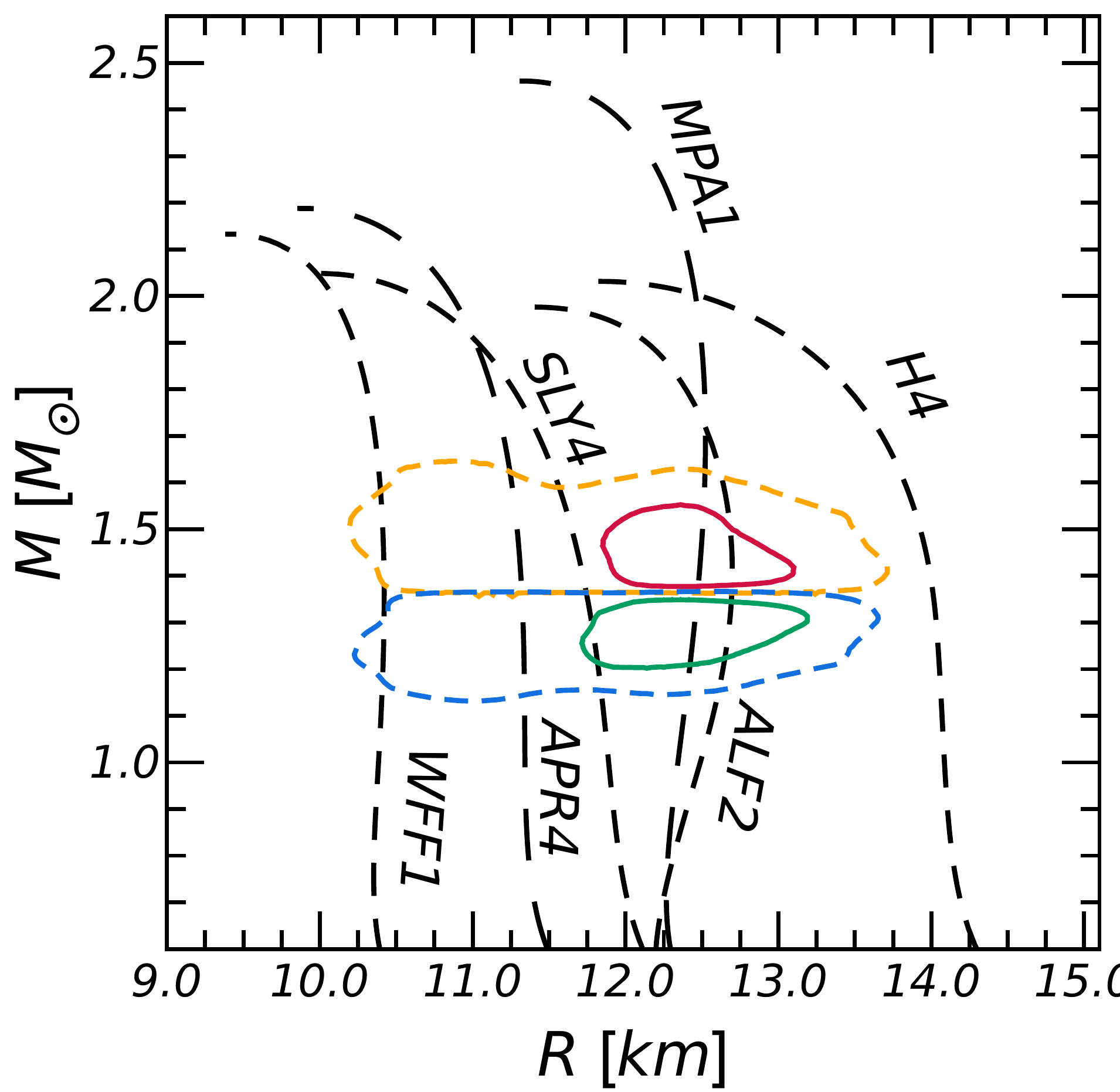} \caption{90\% confidence regions for the posterior 
distribution of mass $M$ and radius $R$ of the two neutron stars of GW170817, inferred using 
our approach (solid contours) and by LVC (dashed contours). Black curves identify the mass-radius 
profiles for some theoretical EoS \cite{Wiringa:1988tp,Alford:2004pf,sly4,Akmal:1998cf,mpa1,Lackey:2005tk}.}
\label{fig:massradius} \end{figure}
Figure~\ref{fig:massradius} shows the joint distribution of mass and radius of the two 
stars of GW170817, together with the confidence intervals derived by LVC \cite{Abbott:2018exr}.
The $M-R$ posteriors derived with our approach fit within the LIGO/Virgo distributions inferred 
through GW data alone. The values of marginalised mass and radius for the two stars, at 90\% 
confidence level are $M_1=1.45^{+0.08}_{-0.06}\ M_{\odot}$, $R_1=12.36^{+0.52}_{-0.38}$ km
and $M_2=1.28^{+0.05}_{-0.06}\ M_{\odot}$, $R_2=12.32^{+0.66}_{-0.43}$ km. While the 
intervals of reconstructed masses are close to the values inferred by LVC, the posterior distributions along the 
$R-$direction obtained by the multi-messenger approach are now effectively narrower.

\begin{figure}[!htbp]
\centering
\includegraphics[width=0.35\textwidth]{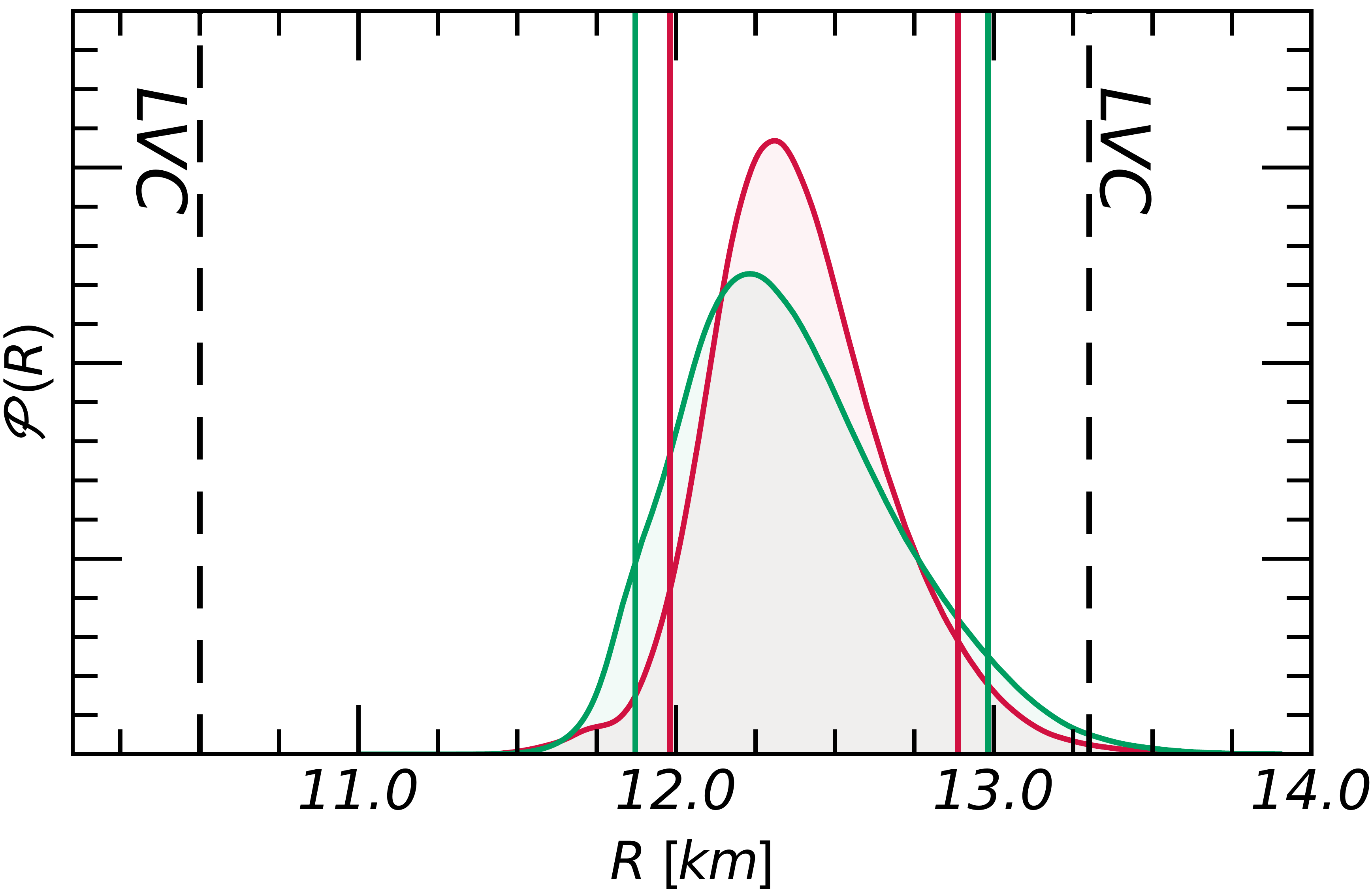}
\caption{Posterior distributions for the radii of the two neutron
stars of GW170817, reconstructed through the
spectral representation. Red and green  colors refer, respectively,
to  $R_1$ and $R_2$. Vertical lines correspond to
intervals at 90\% of probability derived by the LVC (dashed) and
through our analysis (solid).}
\label{fig:radii}
\end{figure}

This is more evident in Fig.~\ref{fig:radii} in which we plot, for the same GW170817 stars, the 
corresponding marginalised distributions of the two radii. The dashed and the solid vertical lines identify the 90\% credible 
interval determined by LVC \cite{Abbott:2018exr}, and by our approach, respectively. As 
already seen in Fig.~\ref{fig:PWposteriors}, the analysis we perform, which  combines GW and 
EM observations, is consistent with the results derived by LIGO/Virgo alone. Most notably, the inclusion 
of the new datasets shrinks the posterior distribution of $R$, which is now determined to an accuracy 
below $10\%$. 
Note that including in our analysis a more massive star, i.e. the NS
with mass and radius estimated through EM observations (specifically
$M_3$ in Table~\ref{tab:Likelihood}), we are probing the EoS in a
region where the energy density is larger with respect to that probed
by the LVC analysis.  

\noindent{\bf{\em Conclusions.}}
Multi-wavelength observations of relativistic sources provide an arena where 
the joint efforts of the astrophysics, high-energy and particle
physics community convey to provide new insights on the fundamental laws of Nature. 
Neutron stars are among the primary targets of this quest, as unique laboratories 
to investigate the behaviour of matter at densities not reproducible
in experiments on Earth. The detection of the first coalescing binary
composed of two neutron stars has allowed the 
LIGO/Virgo collaboration to derive the first GW constraint on the
equation of state of matter in the inner core of neutron stars. A large 
variety of follow-up analyses have been pursued to further exploit the observation of 
the electromagnetic counterpart in coincidence with the gravitational event.
As a result, new bounds have
been derived, which indicate that the EoS of matter in the inner core of a neutron
star is in the soft sector, and produces more compact stars.

In this paper we have made a step forward in this search,
by combining independent measurements of 
NS macroscopic parameters, namely the radius and the tidal deformabilities  from  
the LIGO/Virgo event, and the mass and the radius  derived
from EM observations of low mass X-ray binaries, 
in the spirit of multi-messenger astrophysics. 
Using this approach, we have been able to set tighter constraints on
the radius of the two neutron stars coalescing in GW170817, 
thus supporting, and strengthening,  the observational evidence that neutron star cores 
are composed of soft nuclear matter.

Further detections by ground-based interferometers with higher sensitivities will 
provide new and more accurate observations, which can be used to improve the constraints 
presented in this Letter. It is worth  remarking that
phenomenological parametrizations can introduce systematics 
which may affect the final results of the bayesian analysis. In this regard,  
a wider sample of NS will also allow to test the nature and the relevance of such 
systematics introduced by the specific parametrization \cite{fasano:19}.\\

\begin{acknowledgments}
We thank Giovanni Camelio, Francesco Pannarale and Leonardo Gualtieri
for useful discussions and advices on the methods used in this work. We are 
also indebted with Joonas Nättilä and Coleman Miller for sharing their results on 
the electromagnetic observations which are employed in our analysis.
A.M. acknowledges financial support provided 
under the European Union's H2020 ERC, Starting Grant agreement no. DarkGRA--757480. 
We acknowledge support from the Amaldi Research Center funded by the MIUR program 
``Dipartimento di Eccellenza'' (CUP: B81I18001170001). The authors would like to 
acknowledge networking support by the COST Action CA16104.
\end{acknowledgments}

\bibliography{biblio}

\end{document}